\documentclass[prb,twocolumn,showpacs,amsmath,amssymb]{revtex4}

\usepackage{graphicx}
\usepackage{dcolumn}
\usepackage{bm}

\begin{document}

\title{Small gold clusters on graphene, their mobility and clustering: A DFT study} 

\author{Martin Amft}
\email{martin.amft@fysik.uu.se}
\author{Biplab Sanyal}
\author{Olle Eriksson}
\author{Natalia V. Skorodumova}%
\affiliation{%
Department of Physics and Astronomy, Uppsala University, Box 516,
S-751 20 Uppsala, Sweden}%

\date{\today}%

\begin{abstract}
Motivated by the experimentally observed high mobility of gold atoms on graphene and their tendency to form nanometer-sized clusters, we present a density functional theory study of the ground state structures of small gold clusters on graphene, their mobility and clustering.
Our detailed analysis of the electronic structures identifies the opportunity to form strong gold-gold bonds and the graphene mediated interaction of the pre-adsorbed fragments as the driving forces behind gold's tendency to aggregate on graphene.
While clusters containing up to three gold atoms have one unambiguous ground state structure, both gas phase isomers of a cluster with four gold atoms can be found on graphene.
In the gas phase the diamond shaped Au$_{4}^{D}$ cluster is the ground state structure, whereas the Y shaped Au$_{4}^{Y}$ becomes the actual ground state when adsorbed on graphene.
As we show, both clusters can be produced on graphene by two distinct clustering processes.
We also studied in detail the stepwise formation of a gold dimer out of two pre-adsorbed adatoms, as well as the formation of Au$_{3}$.
All reactions are exothermic and no further activation barriers, apart from the diffusion barriers, were found.
The  diffusion barriers of all studied clusters range from 4 to 36 meV, only, and are substantially exceeded by the adsorption energies of -0.1 to -0.59 eV.
This explains the high mobility of Au$_{1-4}$ on graphene along the C-C bonds.
\end{abstract}


\maketitle

\section{Introduction}
\label{sec:intro}
Historically, carbon allotropes, e.g. graphite, carbon nanotubes, and fullerene have been extensively used for studying the absorption processes of finite-sized particles on them. 
Small coinage metal clusters on single layers of carbon, i.e. graphene, were first used to model the adsorption on graphite surfaces or single-walled carbon nanotubes. \cite{Duffy:1998p10932,Wang:2003p10933,Maiti:2004p10949} 
After the experimental evidence for the existence of graphene was found by Novoselov et al. \cite{Novoselov:2005p11083,Novoselov:2005p11080}, now even awarded with the Nobel Prize in Physics, the study of cluster adsorption on graphene became important in its own right.

Graphene stands as an extraordinary material that offers enormous possibilities for applications in electronics, sensors, biodevices, catalysis, and energy storage. \cite{Geim:2007p6485,Katsnelson:2007p11097,CastroNeto:2009p10058,Geim:2009p11096} 
The unique electronic structure of graphene with a linear dispersion of the electronic structure at Dirac points not only plays an important role in charge transport but in verifying theoretical predictions in quantum electrodynamics by table-top experiments, too. 
As graphene is a zero band gap material, one of the routes to make use of graphene in electronics industry is to create a small band gap by means of functionalization with external chemical agents. \cite{Schedin:2007p11098,Sanyal:2009p11099,Wehling:2008p11088}

By means of density functional theory, the effect of Au nanoparticles on the electronic structure of graphene has been studied.\cite{Carara:2009p11100}
The authors showed that Au$_{38}$ nanoparticles covered with methylthiolate molecules introduce new Dirac-type points due to charge transfer but keeping the graphene layer metallic whereas the bare nanoparticle opens up a small band gap at the Dirac point of graphene.
A recent work \cite{McCreary:2010p10867} has emphasized the effects of the distribution of Au atoms on the charge carrier mobility of graphene. 
They have concluded that the formation of Au clusters increases the mobility whereas a homogeneous distribution reduces the mobility. 

The adsorption and diffusion of gold adatoms on highly oriented pyrolytic graphite (HOPG) have been studied for decades, first, by means of the desorption flux from the surface\cite{ARTHUR:1973p10952}, later on directly with scanning tunneling microscopy \cite{GANZ:1989p10934}, and transmission electron microscopy.\cite{Anton:1998p10931}
Also \textit{ab-initio} calculations on the adsorption of small gold clusters and the diffusion of adatoms and dimers on HOPG have been reported. \cite{Wang:2004p10870,Jensen:2004p10871,Wang:2005p10431}

Very recently the growth of gold nanoparticles on few layers of graphene by physical vapor deposition has been monitored.\cite{Luo:2010p11087} 
The growth process stopped at an average particle size of 6.46$\pm$0.68 nm on a single graphene sheet. 
As atomic-scale manipulation is nowadays possible in experiments, the study of small clusters and their effects on the electronic properties is highly relevant for the future realizations of nanoscale devices. 

A systematic theoretical study of the size variation of small Au clusters adsorbed on graphene is missing. 
Here we address this issue by means of density functional theory (DFT). 
We study the adsorption of Au$_{1-4}$ on graphene, focusing on an analysis of the electronic structures. 
Diffusion barriers are calculated from the total energies and confirm the high mobility of Au$_{1-4}$. 
We study in detail the step-wise formation of a gold dimer out of two pre-adsorbed adatoms, as well as the formation of Au$_{3,4}$.
We show how both Au$_{4}$ isomers from the gas phase can be produced on graphene by two distinct clustering processes.

This paper is organized as follows. 
In the next section \ref{sec:comp} we summarize the computational details of this work. 
In Sec. \ref{sec:GS} the ground state structures of Au$_{1-4}$/graphene, their adsorption energies, charge transfers, and electronic structures are discussed.
Starting from these structures the mobility, i.e. diffusion barriers, and clustering processes have been studied and are presented in Sec. \ref{sec:mobility}.
Finally, Sec. \ref{sec:summ} summarizes and concludes the paper.

\section{Computational details}
\label{sec:comp}
The scalar-relativistic  \textit{ab-initio} DFT calculations were performed using the projector augmented wave (PAW)\cite{Blochl:1994p10844,Kresse:1999p10843} method as implemented in \textsc{vasp}. \cite{Kresse:1996p6093,Kresse:1996p6092}
The exchange-correlation interaction was treated in the generalized gradient approximation (GGA) in the parameterization of Perdew, Burke, and Ernzerhof (PBE). \cite{PERDEW:1996p6520}
A cut-off energy of 600 eV was used and a Gaussian smearing with a width of $\sigma$ = 0.05 eV  for the occupation of the electronic levels.
Spin-polarization was taken into account for all calculations.

The graphene sheet was modeled by a $5\times5$ supercell, i.e. 50 carbon atoms, using the calculated C-C bond length of 1.42~\AA. 
The repeated sheets were separated from each other by 20~\AA\, of vacuum.
A Monkhorst-Pack $\Gamma$-centered $5\times5\times1$ k-point mesh (13 k-points in the irreducible wedge of the Brillouin-Zone) was used for the structural relaxations.
The relaxation cycle was stopped when the Hellman-Feynman forces had become smaller than $5 \cdot 10^{-3}$ eV/{\AA}. 
We also used a finer k-point mesh, i.e. $16\times16\times1$ k-point mesh (130 k-points) and found the changes in the geometry of the systems to be negligible. 
Therefore we decided to use the relaxed structures obtained with the $5\times5\times1$ k-point mesh for total energy ($E_{0}$) and density of states (DOS) calculations.
However we used a finer k-point mesh, i.e. $20\times20\times1$ with 202 k-points, to accurately calculate $E_{0}$ and the DOS of these structures.
Note that cluster adsorption energies, $E_{\mathrm{ads}} = E_{0}(\text{Au$_{1-4}$/graphene}) -  E_{0}(\text{Au$_{1-4}$}) -  E_{0}(\text{graphene})$, are negative when the adsorption is exothermic.
The projected DOS were calculated within Wigner-Seitz spheres of radii 0.863~\AA\, (carbon) and 1.503~\AA\, (gold), using for illustrative purposes a higher Gaussian smearing of 0.1 eV.

In order to obtain the diffusion barriers the total energies of the gold clusters on different binding sites were calculated.
In the calculations the x-y coordinates of the gold atom binding to carbon, as well as the carbon atoms in the rim, were fixed, while the rest of the structure could fully relax.
Since the calculated diffusion barriers are less than 36 meV, we checked their dependence on the used GGA exchange-correlation functional.
Compared to PW91\cite{PERDEW:1992p6521} and RPBE\cite{Hammer:1999p9332}, we found negligible differences only.

The charge distributions and transfers were analyzed by means of the Bader analysis. \cite{Tang:2009p9327} 

\section{Ground state properties of A\lowercase{u}$_{1-4}$/graphene}
\label{sec:GS}
There are different ways to obtain clusters of a certain size on a substrate. 
Here we consider two possibilities: first, mass selected clusters soft-landed onto graphene and, second, the clustering of smaller fragments, already pre-adsorbed on the graphene sheet.

When exploring the first scenario, we obtain the ground state structures of Au$_{1-4}$/graphene, see Fig. \ref{fgr:Figure1} (a)-(d), by fully relaxing the gas phase cluster ground state structures  \cite{BravoPerez:1999p10375,Hakkinen:2000p6499,Gronbeck:2000p6690} in various ways on graphene.  
As will be shown in Sec. \ref{sec:mobility}, the same structures for Au$_{2-4}$ are formed when  starting from smaller pre-adsorbed fragments, i.e. by the second scenario.

The tetramer shows an interesting peculiarity. 
In the gas phase the diamond shaped isomer, Au$_{4}^{D}$, is the ground state. \cite{BravoPerez:1999p10375,Hakkinen:2000p6499,Gronbeck:2000p6690}
According to our calculations, it is 43 meV lower in energy than the Y shaped isomer, Au$_{4}^{Y}$, cf. Refs [\onlinecite{BravoPerez:1999p10375,Gronbeck:2000p6690}].
On graphene the situation is reversed: Au$_{4}^{Y}$/graphene is 137 meV lower in energy than Au$_{4}^{D}$/graphene.
The overall geometry of both isomers is conserved on graphene, cf. Fig. \ref{fgr:Figure1} (d) and (e).
In addition to soft-landing, both isomers can be formed on graphene by clustering of smaller pre-adsorbed fragments, i.e. Au$_{1}$ + Au$_{3}$ $\rightarrow$ Au$_{4}^{D}$ and 2 $\cdot$ Au$_{2}$ $\rightarrow$ Au$_{4}^{Y}$, see Sec. \ref{sec:mobility} for more details.
Note that two of the bonds in Au$_{4}^{Y}$ are comparable to the inter-dimer bond, i.e. 2.55 and 2.6~\AA\, compared to 2.67 and 2.72~\AA, cf. Figs \ref{fgr:Figure1} (b) and (e).

\subsection{Binding sites and energies, charge redistribution}
\label{sec:Eads}

\begin{figure}[hbt]
\includegraphics[width=0.95\linewidth]{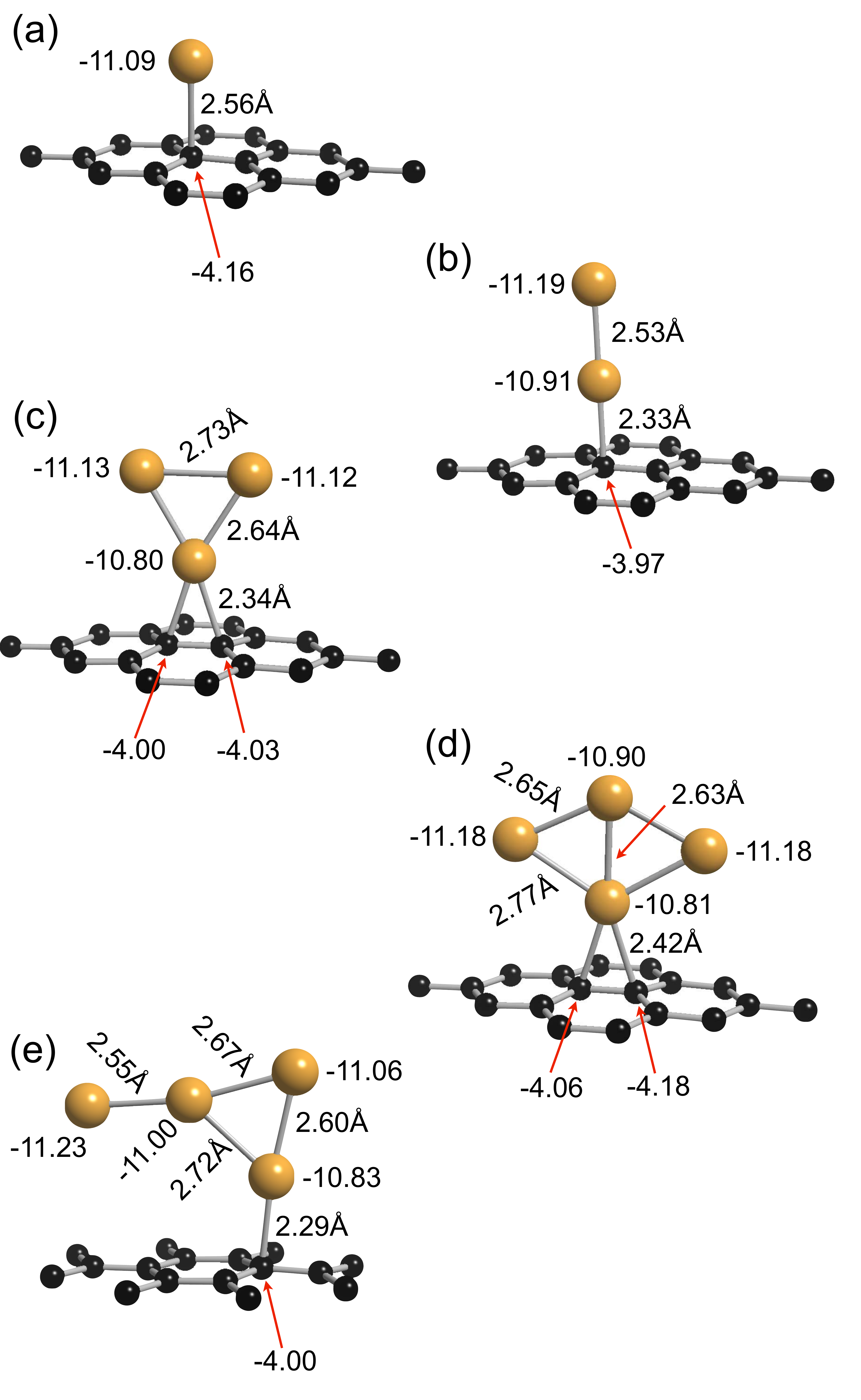}  
 \caption{(Color online) Ground state structures of Au$_{1-4}$ on a 5$\times$5 graphene sheet. Structures (a)-(d) were obtained by soft-landing the clusters from the gas phase. The isomer Au$_{4}^{Y}$ (e) is formed by clustering two pre-adsorbed dimers. It is 137 meV lower in energy than Au$_{4}^{D}$/graphene (d). Also given are relevant bond lengths and Bader charges. Note: in our calculations neutral Au atoms have a Bader charge of -11 and C atoms of -4, respectively.}
\label{fgr:Figure1}
\end{figure}

The adatom, dimer, and Y shaped tetramer prefer the binding site on top of a carbon atom, whereas the trimer and diamond shaped tetramer bind right above a carbon-carbon bond, see Fig. \ref{fgr:Figure1} for an illustration.
It is interesting to observe a trend to minimize the number of gold atoms binding to graphene for all five clusters, i.e. only one Au atom is close to the graphene layer.
This already indicates that the Au-Au interaction is stronger than Au-C interaction, which will be discussed in greater detail below.

The ground state structures of Au$_{1,2}$/graphene, see Fig. \ref{fgr:Figure1} (a) and (b), resemble the adsorption behavior of Au$_{1,2}$ on MgO and graphite.\cite{Amft:2010p11084,Wang:2004p10870,Jalkanen:2007p10951}
The adsorption behavior of the trimer and tetramers on graphene, Figs \ref{fgr:Figure1} (c)-(e), is in contrast to the adsorption of small gold clusters on MgO terraces. 
On MgO the gold trimer and tetramer form two bonds between gold and surface oxygen.
Furthermore only the diamond shaped Au$_{4}$ isomer is stable on that substrate.\cite{Amft:2010p11084}

Care must be taken when comparing our results with earlier studies of Au$_{1-4}$ on HOPG. \cite{Wang:2004p10870,Jensen:2004p10871,Wang:2005p10431} 
First of all the authors relied on the use of the local density approximation (LDA), which only  coincidently describes the bulk properties of graphite better than GGA.
Secondly, the number of k-points was eight at best, making a meaningful calculation of, for instance, the DOS impossible.
Last but not least, spin-polarization was neglected in Refs [\onlinecite{Wang:2004p10870,Wang:2005p10431}].
We find spin-polarization especially important for open shell systems like Au$_{1,3}$/graphene.
They have two distinct peaks, for spin up and down states, respectively, close to the Fermi level, cf. Figs \ref{fgr:Figure4} and \ref{fgr:Figure5}.
Albeit all these methodological differences one can conclude that the ground state structures of Au$_{1-3}$ adsorbed on HOPG and graphene seem to by very similar.
Note that Au$_{3,4}$ have not been studied on graphene before.
The second isomer, Au$_{4}^{Y}$ Fig. \ref{fgr:Figure1} (e), has not even been considered on graphite before.
Note that the tetrahedral structure of Au$_{4}$ is known to be unstable in the gas phase\cite{Gronbeck:2000p6690,Xiao:2004p6455} and neither did we find this structure on graphene.  

\begin{figure}[hbt]
\includegraphics[width=0.95\linewidth]{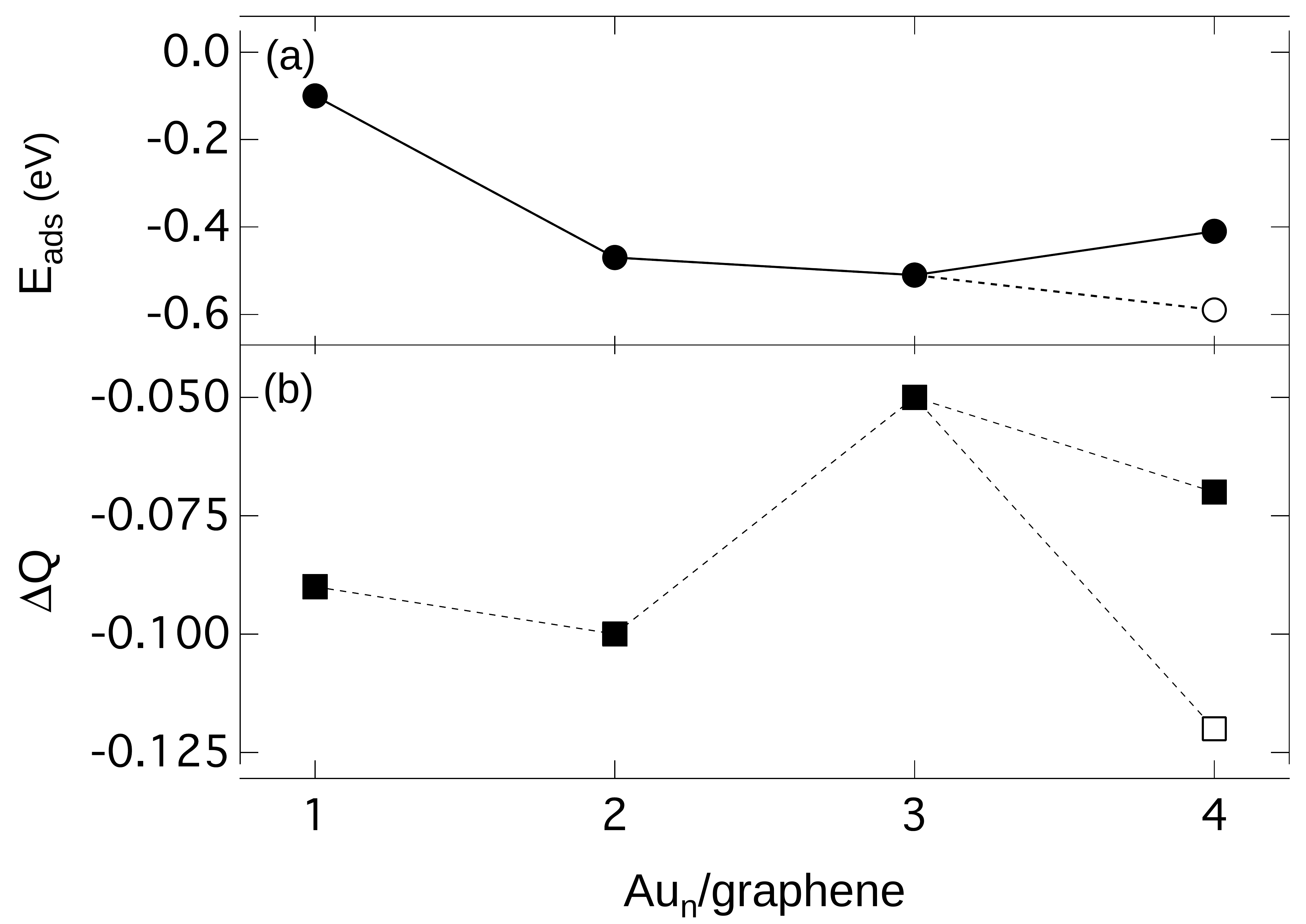}  
 \caption{(Color online) Upper panel (a): Cluster adsorption energy of Au$_{1-4}$ on graphene. Lower panel (b): total charge transferred from the graphene sheet into the clusters, calculated by the Bader analysis. Full symbols: gas phase structures adsorbed on graphene, cf. Figs \ref{fgr:Figure1} (a) - (d). Open symbols: Y shaped Au$_{4}^{Y}$/graphene, cf. Fig. \ref{fgr:Figure1} (e).}
\label{fgr:Figure2}
\end{figure}

Figure \ref{fgr:Figure2} (a) shows the cluster adsorption energies $E_{\mathrm{ads}}$ of the Au$_{1-3}$ ground state structures, as well as of both Au$_{4}$ isomers. 
Au$_{4}^{Y}$ has the highest $E_{\mathrm{ads}}$ of -0.59 eV and the adatom having with -0.1 eV the lowest.
As mentioned above, Au$_{4}^{Y}$ is not only the ground state structure on graphene, but also adsorbs stronger on it than Au$_{4}^{D}$, i.e. $E_{\mathrm{ads}} =$ -0.59 vs. -0.41 eV, respectively.

In all the five cases the whole gold cluster receives, according to our Bader analysis, a small charge of up to -0.12 $e^{-}$ from graphene, see Fig. \ref{fgr:Figure2} (b).
Hence adsorbing Au$_{1-4}$ on graphene corresponds to a p-doping of the material.
A clear odd-even oscillation, depending on the number of gold atoms in the cluster, can be seen in the charge transfer.
It is generally known that the electronic characteristics of low-dimensional gold structures can exhibit odd/even oscillations depending on the parity of the number of Au atoms,
which are determined by the opening/closing of the $s$-shell.\cite{Skorodumova:2005p10022,Grigoriev:2006p10023,Amft:2010p11084}

In addition to the charge transfer into the clusters, there occurs also a significant charge redistribution within the clusters themselves, see Bader charges of the individual atoms in Fig. \ref{fgr:Figure1}.
The insets in Figs \ref{fgr:Figure4} - \ref{fgr:Figure8} also illustrate it by showing isosurfaces of the charge density redistribution, i.e. $\Delta\rho = \rho(\text{Au}_{n}/\text{graphene}) - \rho(\text{Au}_{n}) - \rho(\text{graphene})$, upon the cluster adsorption on graphene.
In the case of Au$_{2-4}$ up to 0.19 electrons are redistributed from the gold atom binding to graphene into the rest of the cluster.

Concerning the adsorption strength of these small gold clusters on graphene we can summarize that two effects exist, which can compete with each other. 
First, the attraction due to the formation of chemical bonds between gold and carbon.
Second, there also exists a Coulomb interaction between the charged gold atoms in the cluster and the carbon atoms in the proximity of the adsorption site.
Mostly this Coulomb interaction is repulsive in nature, as it can be seen from the extra charge in the clusters that tends to be concentrated as far away from the graphene sheet as possible, see Figs \ref{fgr:Figure1} (a) - (c).

Although the extra charge density concentrates on the two outer atoms in Au$_{4}^{D}$/graphene, those two atoms are repelled away from the graphene sheet, recognizable from the 5\% stretched bonds with the bottom gold atom, see Fig. \ref{fgr:Figure1} (d).
In contrast to these four systems, the Coulomb interaction between the adsorbed Au$_{4}^{Y}$ and the underlying graphene sheet is predominately attractive, i.e. between the left-most gold atom, cf. Fig. \ref{fgr:Figure1} (e), carrying most of the additional charge, and the positively charged carbon atoms under it, cf. inset in Fig. \ref{fgr:Figure7}.

The comparatively small adsorption energy of the single gold atom is also explained by these competing effects, see Fig. \ref{fgr:Figure1} (a), which is manifested in the 5-9 \% longer Au-C distance of Au$_{1}$/graphene compared to the other clusters.

\begin{figure}[hbt]
\includegraphics[width=0.95\linewidth]{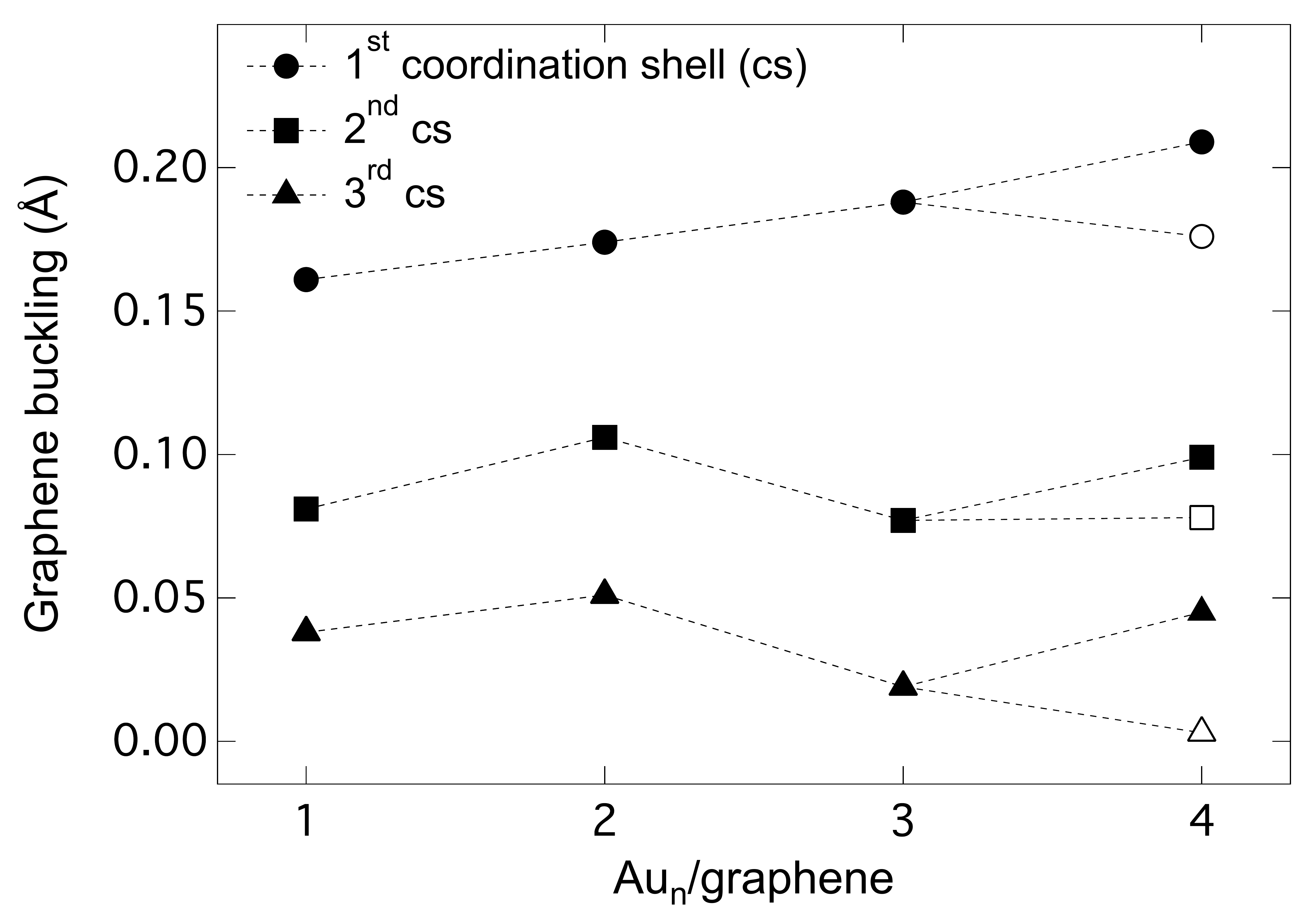}  
 \caption{(Color online) Buckling, i.e. vertical distortion, of the first three coordination shells around the adsorption sites of Au$_{1-4}$. Full symbols: gas phase structures adsorbed on graphene, cf.  Figs \ref{fgr:Figure1} (a) - (d). Open symbols: Au$_{4}^{Y}$/graphene, cf. Fig. \ref{fgr:Figure1} (e).}
\label{fgr:Figure3}
\end{figure}

It has, predominantly qualitatively, been reported in the literature that the adsorption of Au$_{1-4}$ on HOPG\cite{Wang:2004p10870,Jensen:2004p10871,Wang:2005p10431,Akola:2006p10034} and of Au$_{1,2}$ on graphene\cite{Chan:2008p10033,Varns:2008p10877} leads to a shift of the carbon atoms in the vicinity of the adsorption site towards the gold.
Due to the strong $sp^{2}$ interatomic C-C bonds the vertical distortion is not solely restricted to the carbon atoms underneath gold.
Instead it spreads at least to the third coordination shell around the adsorbate, cf. Fig. \ref{fgr:Figure3}.
Lets consider for the moment only the gas phase structures adsorbed on graphene. 
The full symbols in Fig. \ref{fgr:Figure3} show that the vertical distortion of the carbon atoms binding to the Au atom scales approximately linear with the size of the gold cluster.
On the other hand, the buckling in the second and third coordination shell does show a clear odd-even oscillation depending on the number of Au atoms,  Fig. \ref{fgr:Figure3} (full symbols).
But the actual tetramer ground state geometry Au$_{4}^{Y}$/graphene breaks these trends, showing a significantly smaller distortion of the C atom positions than Au$_{4}^{D}$/graphene, see Fig. \ref{fgr:Figure3} (open symbols).

Along with the buckling comes a polarization of the graphene sheet, i.e. charge deviations from otherwise neutral charge state of the carbon atoms, of up to $\pm$ 0.18 $e^{-}$.
For the systems under consideration the polarization does not show such a clear pattern as observed on HOPG. \cite{Wang:2004p10870,Wang:2005p10431}
For instance, the carbon atom binding to Au$_{1,3}$ and Au$_{4}^{D}$ gain additional electrons, whereas it stays neutral when binding to Au$_{4}^{Y}$ or even looses some charge as under Au$_{2}$.
Also in the second and third coordination shells around the binding sites a general pattern could not be observed, i.e. charge losses and gains could be observed within the same coordination shell.  

\subsection{Electronic structure of Au$_{1,3}$/graphene}
\label{sec:DOS13}
Both the gold adatom and trimer possess an unpaired 6$s$ electron that gives rise to a total spin moment of 1$\mu_{B}$ in the system.

\begin{figure}[hbt]
\includegraphics[width=0.95\linewidth]{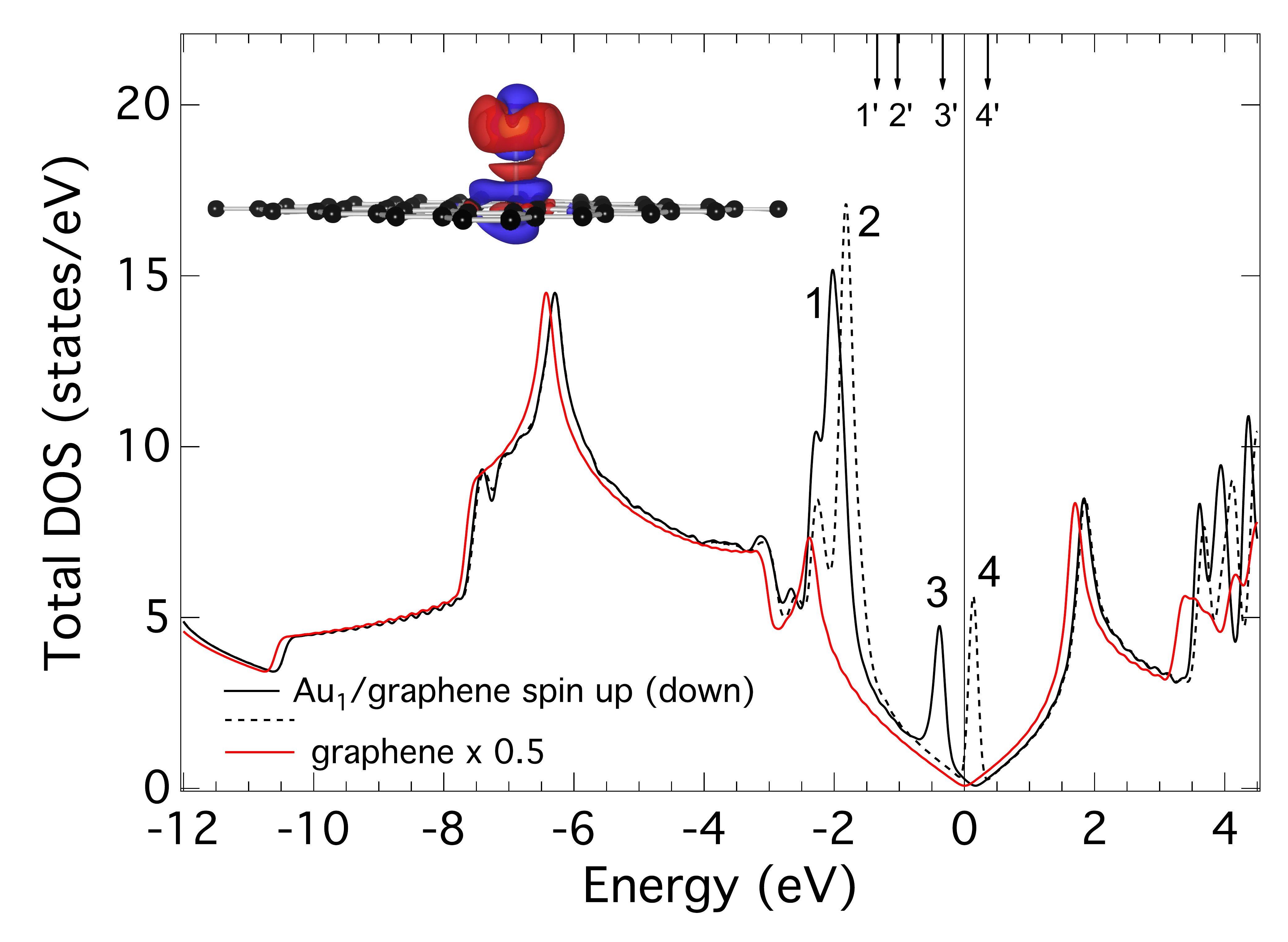}  
 \caption{(Color online) Projected spin-polarized total DOS of Au$_{1}$/graphene (black lines; solid   spin up; dashed spin down) and for comparison pure graphene (red line). Note: the Fermi energies of both systems were shifted to 0 eV. Inset: charge density redistribution due to adsorption of Au$_{1}$ on graphene;  increase (decrease) of the charge density in red (blue). The shown isosurfaces correspond to $\pm$ 7$\cdot$10$^{-4}$ e$^{-}$/\AA$^{3}$. The arrows on the top show the atomic energy levels of a single gold atoms.}
\label{fgr:Figure4}
\end{figure}

\begin{figure}[hbt]
\includegraphics[width=0.95\linewidth]{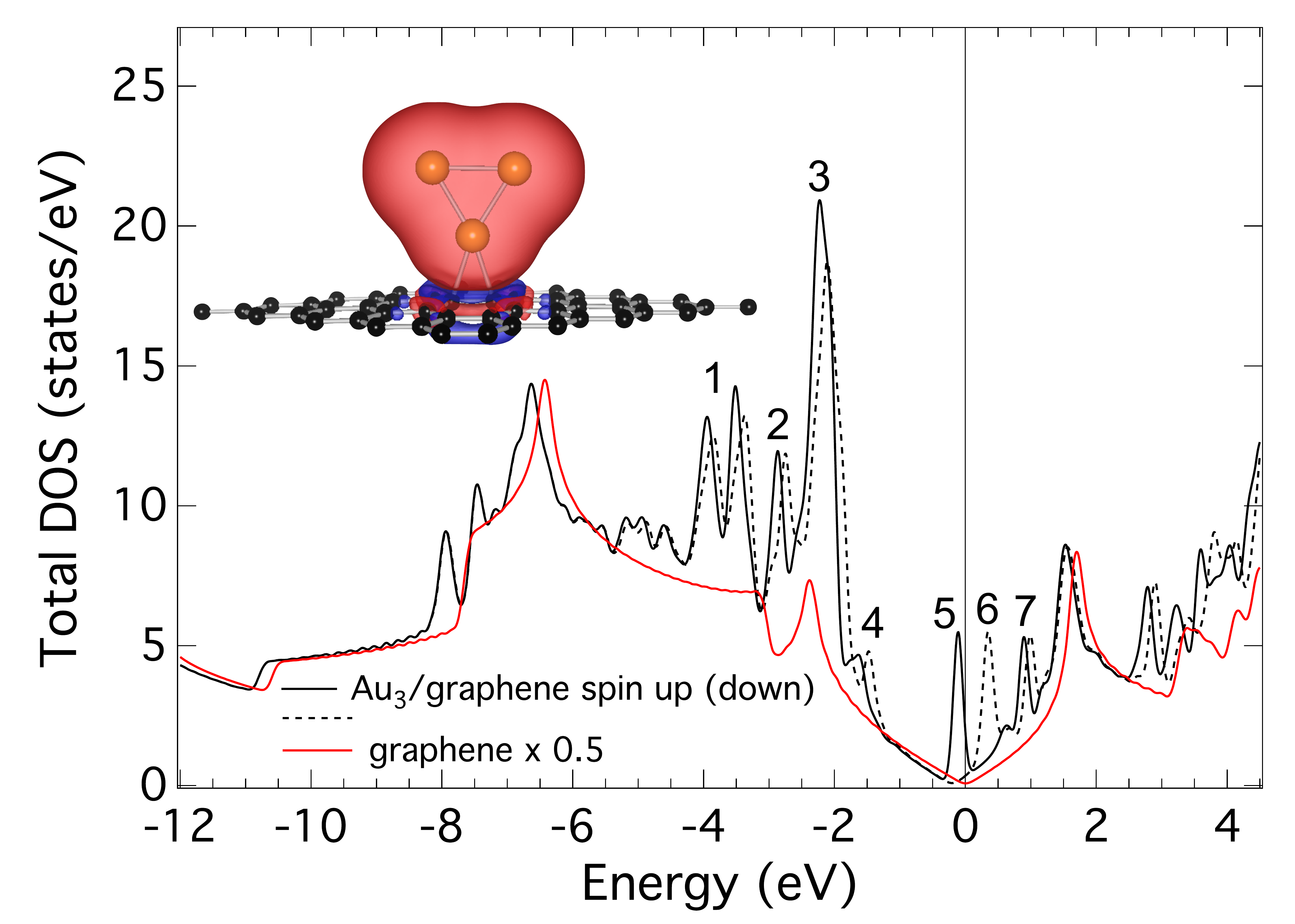}  
 \caption{(Color online) Projected spin-polarized total DOS of Au$_{3}$/graphene (black lines; solid   spin up; dashed spin down) and for comparison pure graphene (red line). Note: the Fermi energies of both systems were shifted to 0 eV. Inset: charge density redistribution due to adsorption of Au$_{3}$ on graphene;  increase (decrease) of the charge density in red (blue). The shown isosurfaces correspond to $\pm$ 7$\cdot$10$^{-4}$ e$^{-}$/\AA$^{3}$.}
\label{fgr:Figure5}
\end{figure}

Figures \ref{fgr:Figure4} and \ref{fgr:Figure5} show the total spin-polarized DOS of Au$_{1,3}$/graphene and, for comparison, also the total DOS of pure graphene.
For illustration the atomic energy levels of an isolated gold atom are shown as arrows in Fig. \ref{fgr:Figure4}.
Only in this simple case the origin of the electronic states of the adsorbed adatom can easily be identified from the gas phase species. 

From the $lm$ and site decomposed DOS (not shown) we identified the character of the labeled structures 1-4 in the DOS of Au$_{1}$/graphene and accordingly for the bigger clusters as well.

In Fig. \ref{fgr:Figure4} the structures no. 1 (2) are gold 5$d$-states with spin polarization up (down).
The higher peaks in 1 and 2 consist partially of $d_{z^{2}}$-states that overlap with $p_{z}$-states of the underlying carbon atom.
Also the gold 6$s$-state near the Fermi level, i.e. peak 3, hybridizes with the $p_{z}$-state to form a $\sigma$-bond.
The unoccupied states in peak 4 are of similar character as those in no. 3.
Compared to the atomic states, i.e. arrows 1' and 2' in Fig. \ref{fgr:Figure4}, the $d$-states are split  upon adsorption of the gold atom on graphene.
Furthermore they are shifted approximately 0.7 eV downwards in energy.
The gap between the highest occupied states and the lowest unoccupied states $\Delta E_{G}$, i.e. between the peaks 3 and 4,  is 0.15 eV narrower than in the gas phase, i.e. 3' and 4'.

For the bigger clusters, we concentrate our efforts on the identification of the most important features of the densities of states, i.e. Figs \ref{fgr:Figure5} - \ref{fgr:Figure8}, and their contribution to the formation of the clusters and their bonding to graphene.

Already at a first glance at the DOS of Au$_{3}$/graphene, Fig. \ref{fgr:Figure5}, one can recognize the higher complexity of the structure.
The two peaks labeled with 1 comprise predominantly of $d_{yz}$ and $d_{x^{2}-y^{2}}$-states of the gold atom closest to carbon, hybridizing with its $p$-states. 
The peaks no. 2 are predominantly  $d_{x^{2}-y^{2}}$-states located at the two gold atoms at the top.
The main features, no. 3, mostly contain $d$-states of all characters from the top gold atoms.
Most of the $d_{z^{2}}$-states of all three gold atoms are concentrated in this peak, forming the chemical bond to carbon by hybridizing with its $p_{z}$-states.
Peak no. 4 consists of intracluster $d_{xz}$-$d_{z^{2}}$ states. 
The peak no. 5 directly below the Fermi energy contains the 6$s$-states of the top gold atoms that interact only indirectly, via $d_{xz}$-states of the gold atom closest to graphene, with the carbon $p_{z}$-states.
Clearly, most of the spin-moment of the whole system is concentrated in these states.
Finally, the peaks 6 and 7 in the unoccupied DOS are overlaps of empty 6$s$-states and carbon $p_{z}$-states from the top gold atoms and the bottom one, respectively. 

From the shown DOS, especially the peaks near the Fermi energy, one can conclude that even a small external bias of less than $\pm\,$0.5 V will significantly increase the electrical conductivity of these systems.

The charge density redistributions upon adsorption $\Delta \rho$ are shown as insets in Figs \ref{fgr:Figure4} and \ref{fgr:Figure5}.
The earlier mentioned polarization of the carbon atoms in the first three coordination shells around the adsorption sites can be seen. 
The extra charge in Au$_{1}$ also leads to a charge density redistribution within the gold atom, whereas the gold atoms in the trimer seem to equally share the additional charge from the substrate. 
Therefore it is important to take the Bader analysis, Fig. \ref{fgr:Figure1}, into consideration as well that showed a charge depletion of the bottom gold atom.

\subsection{Electronic structure of Au$_{2,4}$/graphene}
\label{sec:DOS24}
Obviously, the total magnetic moment of the closed shell structures Au$_{2,4}$ must be zero in the gas phase.
This is unchanged by adsorbing them on graphene.

\begin{figure}[hbt]
\includegraphics[width=0.95\linewidth]{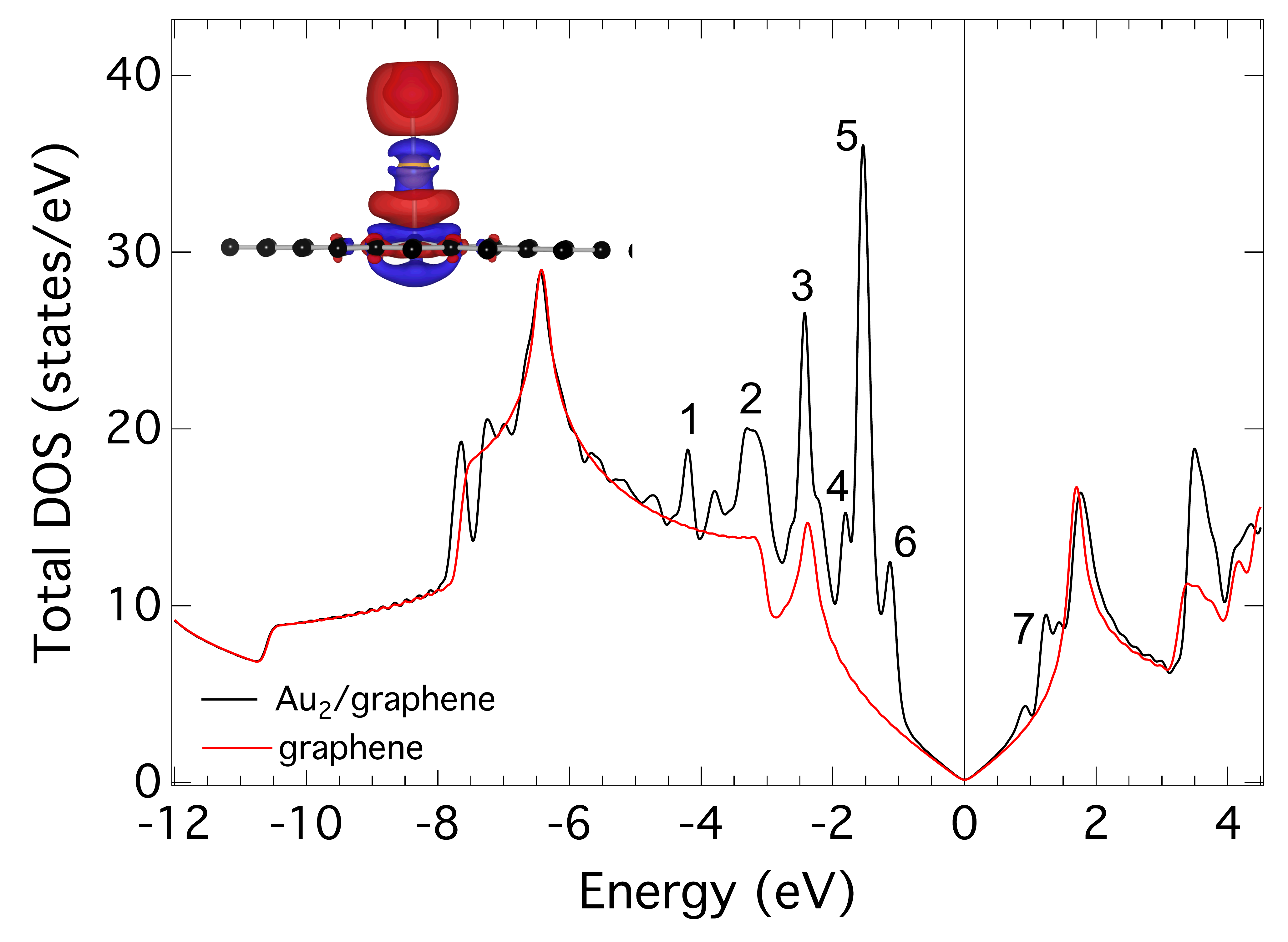}  
 \caption{(Color online) Projected total DOS of Au$_{2}$/graphene (black line) and for comparison pure graphene (red line). Note: the Fermi energies of both systems were shifted to 0 eV. Inset: charge density redistribution due to adsorption of Au$_{2}$ on graphene;  increase (decrease) of the charge density in red (blue). The shown isosurfaces correspond to $\pm$ 7$\cdot$10$^{-4}$ e$^{-}$/\AA$^{3}$.}
\label{fgr:Figure6}
\end{figure}

Upon adsorption the symmetry of the electronic structure of the dimer is broken, see Fig. \ref{fgr:Figure6}.
From the $lm$ and site decomposed DOS (not shown), we see that peak no. 1 consists of $d_{z^{2}}$-states.
The broader feature no. 2 and peak no. 3 are $d_{xy}$, $d_{yz}$, and $d_{x^{2}-y^{2}}$-states, predominantly from the gold atom closest to carbon, hybridizing with carbon $p_{x,y}$-states.
Peaks 4 and 5 do not contribute to the chemical bonding to graphene, as they are $s$-$d$-states  localized at the cluster. 
Whereas no. 6, mostly $s$-$d_{z^{2}}$-states at the top gold atom, indirectly interacts with carbon $p_{z}$ via $d_{z^{2}}$ at the bottom gold atom.
Finally, the two peaks at no. 7 are predominantly unoccupied 6$s$-states of the gold atom close to carbon, overlapping empty $p_{z}$-states of carbon.
Note also that the states in peaks 4-7 substantially narrow the cone in graphene's DOS around the Fermi energy, leaving it only undisturbed in the interval $E_{F}\,\pm\,$0.5 eV, Fig. \ref{fgr:Figure6}.

\begin{figure}[hbt]
\includegraphics[width=0.95\linewidth]{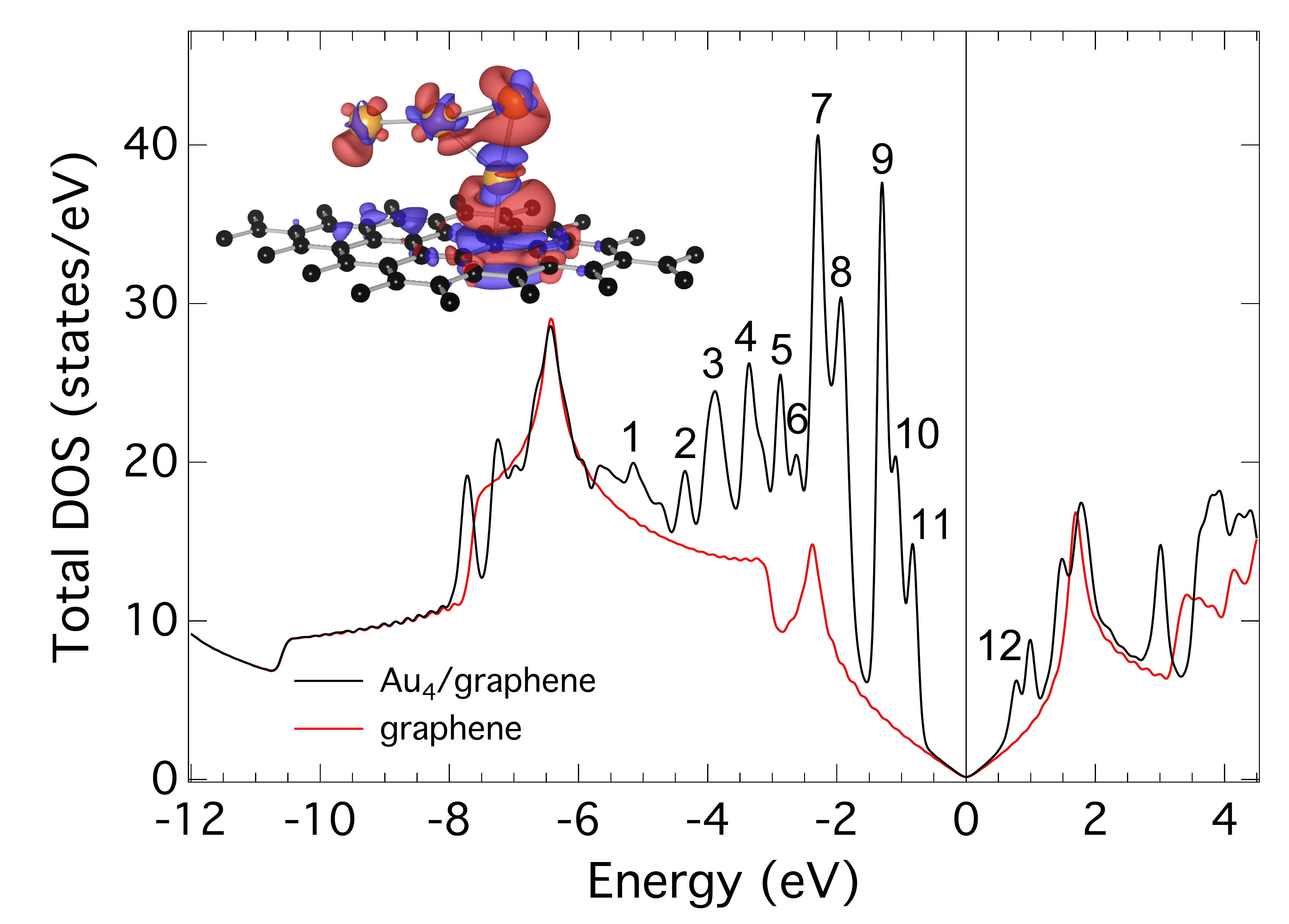}  
 \caption{(Color online) Projected total DOS of the Y shaped Au$_{4}$/graphene (black line), see  Fig. \ref{fgr:Figure1} (e), and for comparison pure graphene (red line). Note: the Fermi energies of both systems were shifted to 0 eV. Inset: charge density redistribution due to adsorption of Au$_{4}$ on graphene;  increase (decrease) of the charge density in red (blue). The shown isosurfaces correspond to $\pm$ 7$\cdot$10$^{-4}$ e$^{-}$/\AA$^{3}$.}
\label{fgr:Figure7}
\end{figure}

To summarize the predominant nature of the labeled peaks in the DOS of Au$_{4}^{Y}$/graphene, Fig. \ref{fgr:Figure7}, we label the gold atoms in the following way, see Fig. \ref{fgr:Figure1} (e): Au$^{l}$ (left), Au$^{m}$ (middle), Au$^{t}$ (top), and Au$^{b}$ (bottom). 
The carbon atom under Au$^{b}$, forming the chemical bond with the cluster, is named C$^{b}$.

Peaks no. 1 and 2 consist of $s$-$d$-states, $d_{z^{2}}$-states respectively, localized on the cluster only.
Peak no. 3 consists of $d$-states that contribute to the chemical bond by overlapping with $p_{y}$-states of C$^{b}$, i.e. $d_{xz}$-states on Au$^{m}$ and $d_{yz}$-states on Au$^{t,b}$.
The peak no. 4 and its shoulder consist mainly of $d_{xy}$ and $d_{x^{2}-y^{2}}$ on Au$^{b}$ overlapping with carbon $p$-states as well.
Peak no. 5 are $d_{x^{2}-y^{2}}$-states solely localized within the cluster, while no. 6 are $s$-$d_{z^{2}}$-states of Au$^{b}$ hybridizing strongly with $p_{z}$ of C$^{b}$.
The highest peak no. 7 consists of $d_{xy}$ and $d_{x^{2}-y^{2}}$-states of Au$^{l}$, Au$^{m}$, and Au$^{t}$, i.e. a major part of the strong gold-gold bonds that hold the cluster together.
Peaks no. 8 and 9 are $d$-states almost exclusively localized on Au$^{m}$ and Au$^{l}$, respectively, as well as peaks no. 10 ($d_{x^{2}-y^{2}}$-states) and 11 ($s$-$d_{xy}$-states) which are localized on Au$^{l}$.
As we saw for the other clusters the unoccupied states at peak no. 12 are $s$-states, in the present case of Au$^{t}$ and Au$^{b}$, overlapping with unoccupied p$_{z}$-states of C$^{b}$.

\begin{figure}[hbt]
\includegraphics[width=0.95\linewidth]{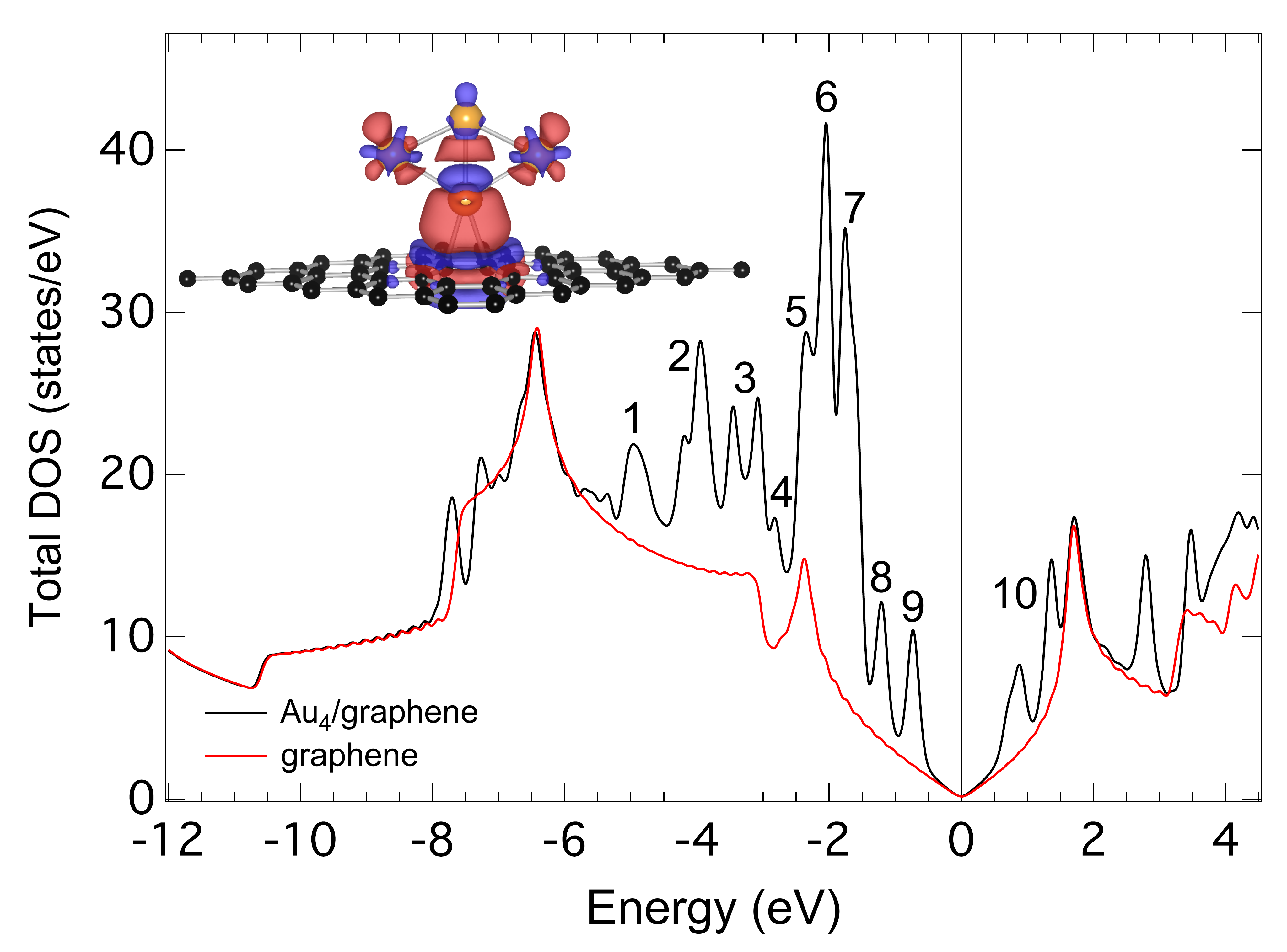}  
 \caption{(Color online) Projected total DOS of the diamond shaped Au$_{4}$/graphene (black line),   see Fig. \ref{fgr:Figure1} (e), and for comparison pure graphene (red line). Note: the Fermi energies of both systems were shifted to 0 eV. Inset: charge density redistribution due to adsorption of Au$_{4}$ on graphene;  increase (decrease) of the charge density in red (blue). The shown isosurfaces correspond to $\pm$ 7$\cdot$10$^{-4}$ e$^{-}$/\AA$^{3}$.}
\label{fgr:Figure8}
\end{figure}

Despite its more symmetric geometry the DOS of Au$_{4}^{D}$/graphene, Fig. \ref{fgr:Figure8}, has many features which can be recognized from the other isomer, cf. Fig. \ref{fgr:Figure7}.
For convenience we label the gold atoms in Au$_{4}^{D}$/graphene again: Au$^{b}$ (bottom), Au$^{m}$ (middle), and Au$^{t}$ (top), see Fig. \ref{fgr:Figure1} (d).
The two carbon atoms, forming bonds with the cluster, have a very similar electronic structure and we therefore simply name them both C$^{b}$.

The relatively broad peak no. 1 consists mainly of $d_{xz}$-states of Au$^{b}$ and Au$^{t}$.
The shoulder below peak no. 2, are $d_{z^{2}}$-$d_{xy}$-cluster states of Au$^{b}$ and Au$^{t}$, with a smaller amount of intermixed  $s$-states that hybridize with the $p$-states of C$^{b}$.
The actual peak no. 2 and the double-peak no. 3 are predominantly $d_{yz}$ and $d_{x^{2}-y^{2}}$-states of Au$^{b}$ and Au$^{t}$, respectively.
The lower peak no. 4 contains $s$-$d_{xy}$-states of Au$^{b}$ that also overlap with C$^{b}$ $p_{z}$-states.
The high peaks no. 5 to 7 are the strong intracluster bonds, formed out of $d$-states. 
Peak no. 8 on the other hand consists mainly of $d_{xz}$-states of Au$^{m}$, as well as $d_{z^{2}}$-states of Au$^{t}$ and Au$^{b}$, contributing to the chemical bond to carbon by hybridizing with $p_{z}$-states of C$^{b}$.
The highest occupied states, peak no. 9, are again intracluster bonds, formed out of $s$-states of the Au$^{m}$ and $d_{xz}$-states of Au$^{t}$ and Au$^{b}$.
As for the other clusters the lowest unoccupied states, peaks no. 10, are unoccupied $s$-states on all the four gold atoms, that overlap with unoccupied C$^{b}$ $p_{z}$-states.

The gap between the highest occupied states and the lowest unoccupied cluster states  $\Delta E_{G}$ in Fig. \ref{fgr:Figure7} and \ref{fgr:Figure8} is very similar, but smaller than that in the dimer case, Fig. \ref{fgr:Figure6}.
Hence, in contrast to Au$_{1,3}$/graphene, an external bias of more than $\pm\,$0.5 V is needed to see a significant increase in the electrical conductivity.

\section{Mobility and clustering}
\label{sec:mobility}

In this section we discuss the mobility and initial clustering processes of gold on graphene.
First experiments on the diffusion and desorption of gold atoms on graphite date almost three decades back.\cite{ARTHUR:1973p10952}
With the rise of STM's quantitative estimates of the diffusion and desorption barriers became feasible.  
These experiments revealed that the binding energies of Au$_{1,2}$ on graphite exceed their diffusion barriers.\cite{GANZ:1989p10934}
Computations of gold adatoms and dimers on HOPG supported these experiments.\cite{Jensen:2004p10871}

\subsection{Diffusion along the C-C bonds}
We determined the diffusion barriers for Au$_{1-4}$ on graphene from the total energies on the top (t) and bridge (b) positions, as well as from one position in between those two (tb).
These energies are sufficient, since none of the five clusters was found to bind to graphene at the hollow sites, i.e. in the center of a carbon hexagon.

\begin{figure}[hbt]
\includegraphics[width=0.95\linewidth]{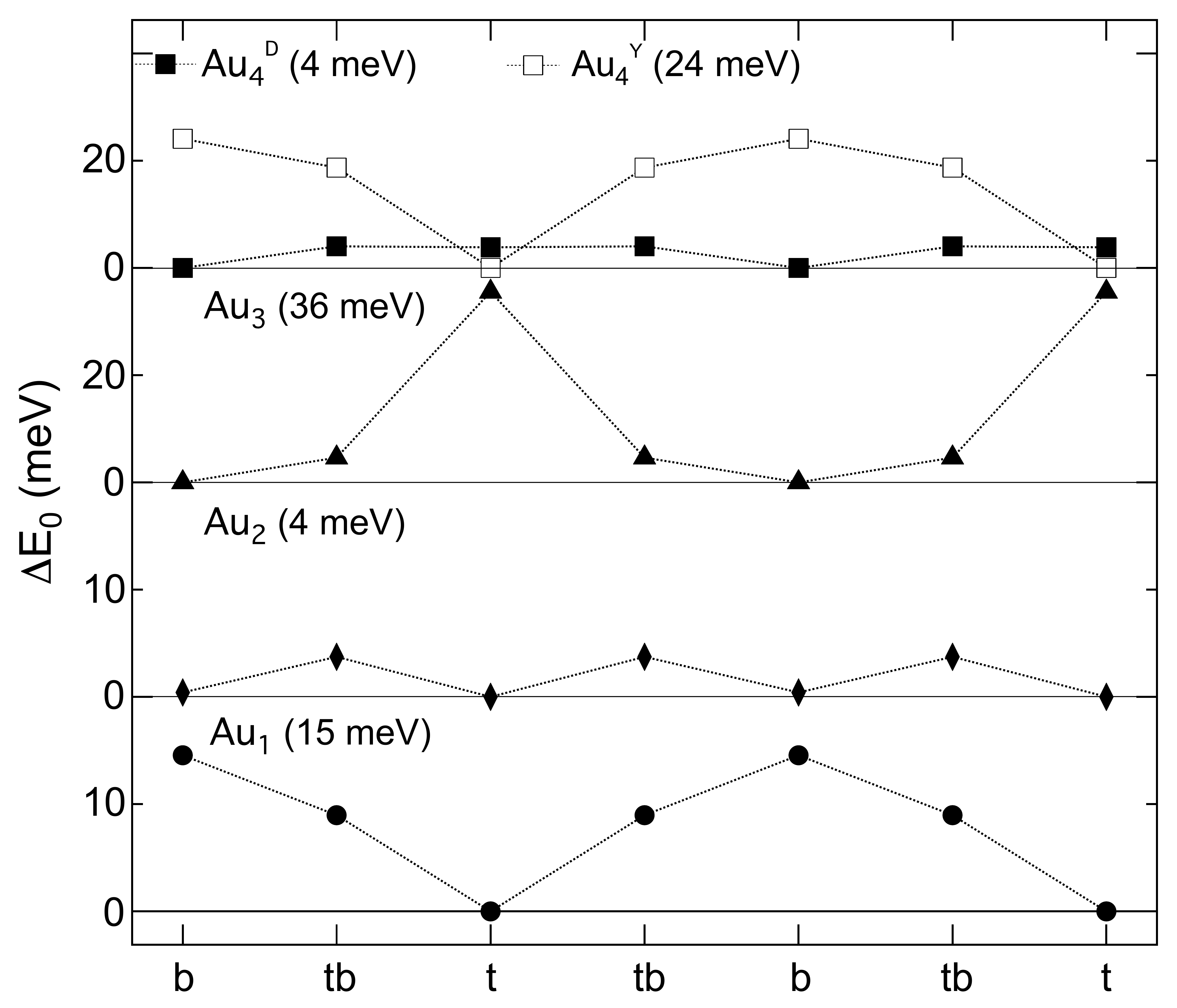}  
 \caption{(Color online) Difference in the total energies $\Delta E_{0}$ at the top (t), bridge (b), and a positions in between (tb) of Au$_{1-3}$ and Au$_{4}^{Y,D}$ on graphene. E$_{0}$ of the respective cluster ground state was subtracted to show the diffusion barriers along the C-C bonds (also given in parenthesis). Note the smaller scales for Au$_{3,4}$.}
\label{fgr:Figure9}
\end{figure}

Figure \ref{fgr:Figure9} shows the differences in the total energies, $\Delta E_{0}$, of the five clusters with respect to the total energy at their binding site, i.e. top (Au$_{1,2}$ and Au$_{4}^{Y}$) and bridge for the remaining two.

Note that only Au$_{2}$ binds on both the top and bridge position, while the other four clusters have a saddle point at the bridge (Au$_{1}$ and Au$_{4}^{Y}$) or top (Au$_{3}$ and Au$_{4}^{D}$) position, respectively, i.e. if allowed they relax to their stable binding site. 
These transition states define the diffusion barriers of the gold clusters along the C-C bonds.
For convenience the diffusion barriers are also explicitly given in Fig. \ref{fgr:Figure9}.
They range from only 4 meV for Au$_{2}$ and Au$_{4}^{D}$, making them highly mobile, to 36 meV for the least mobile trimer.

The adsorption energies of all five gold clusters on graphene, cf. Fig. \ref{fgr:Figure2} (a), exceed their diffusion barriers by at least one order of magnitude.
Hence, while these small gold clusters will rather strongly adsorb onto the graphene sheet, they will easily diffuse even at very low temperatures. 


\subsection{Clustering of pre-adsorbed fragments}

\begin{figure}[hbt]
\includegraphics[width=0.95\linewidth]{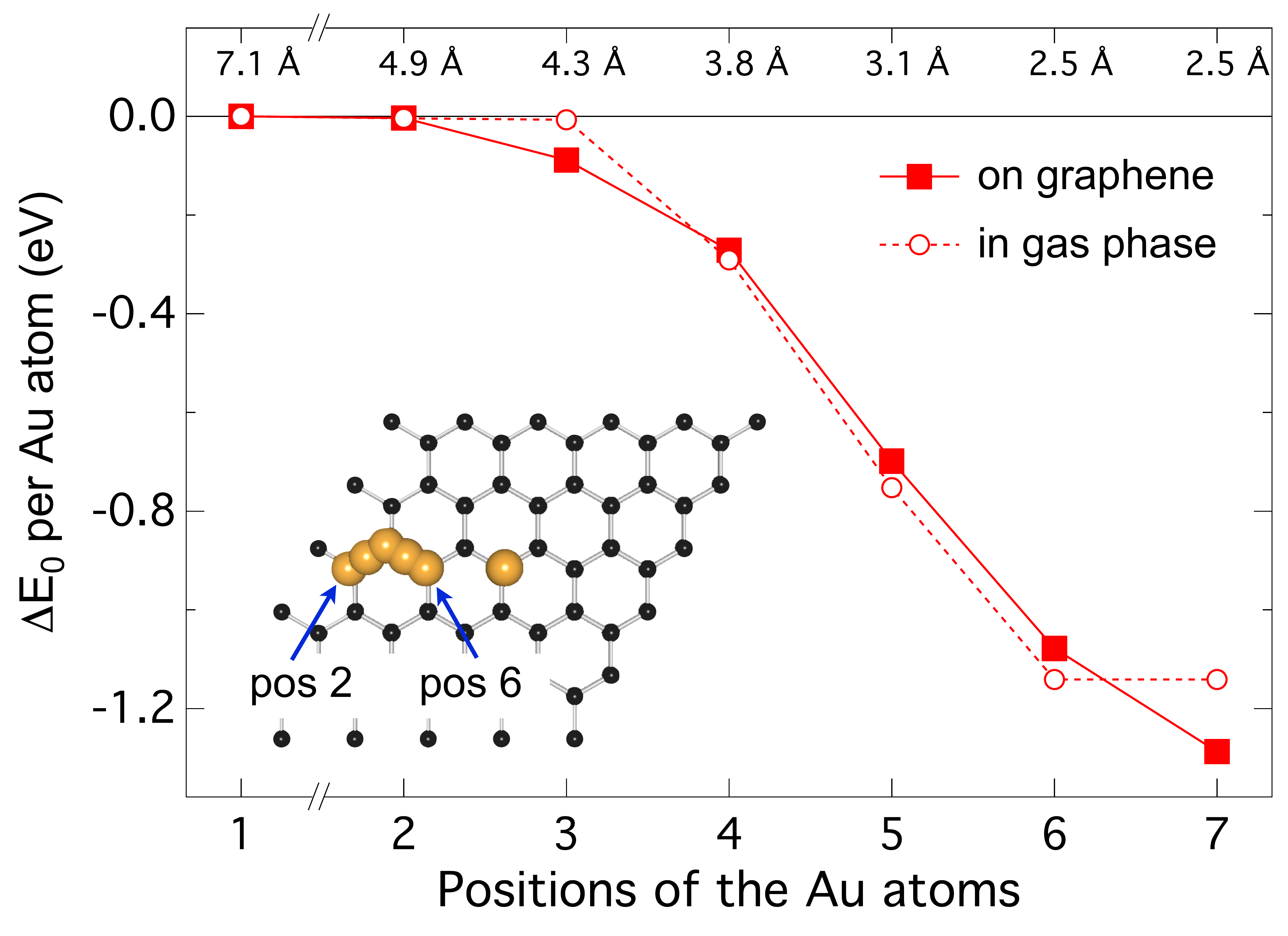}  
 \caption{(Color online) Changes in the total energy $\Delta E_{0}$ when forming Au$_{2}$/graphene from two pre-adsorbed adatoms (solid symbols) and for the dimer formation in the gas phase (open symbols). The x-y coordinates of the Au adatoms were kept fixed for positions 1-6. The system relaxes without an additional activation barrier from position 6 to 7, the dimer ground state structure (Fig. \ref{fgr:Figure2} (b)). The Au-Au distances for the different positions are given on the top axis.}
\label{fgr:Figure10}
\end{figure}

\begin{figure}[hbt]
\includegraphics[width=0.95\linewidth]{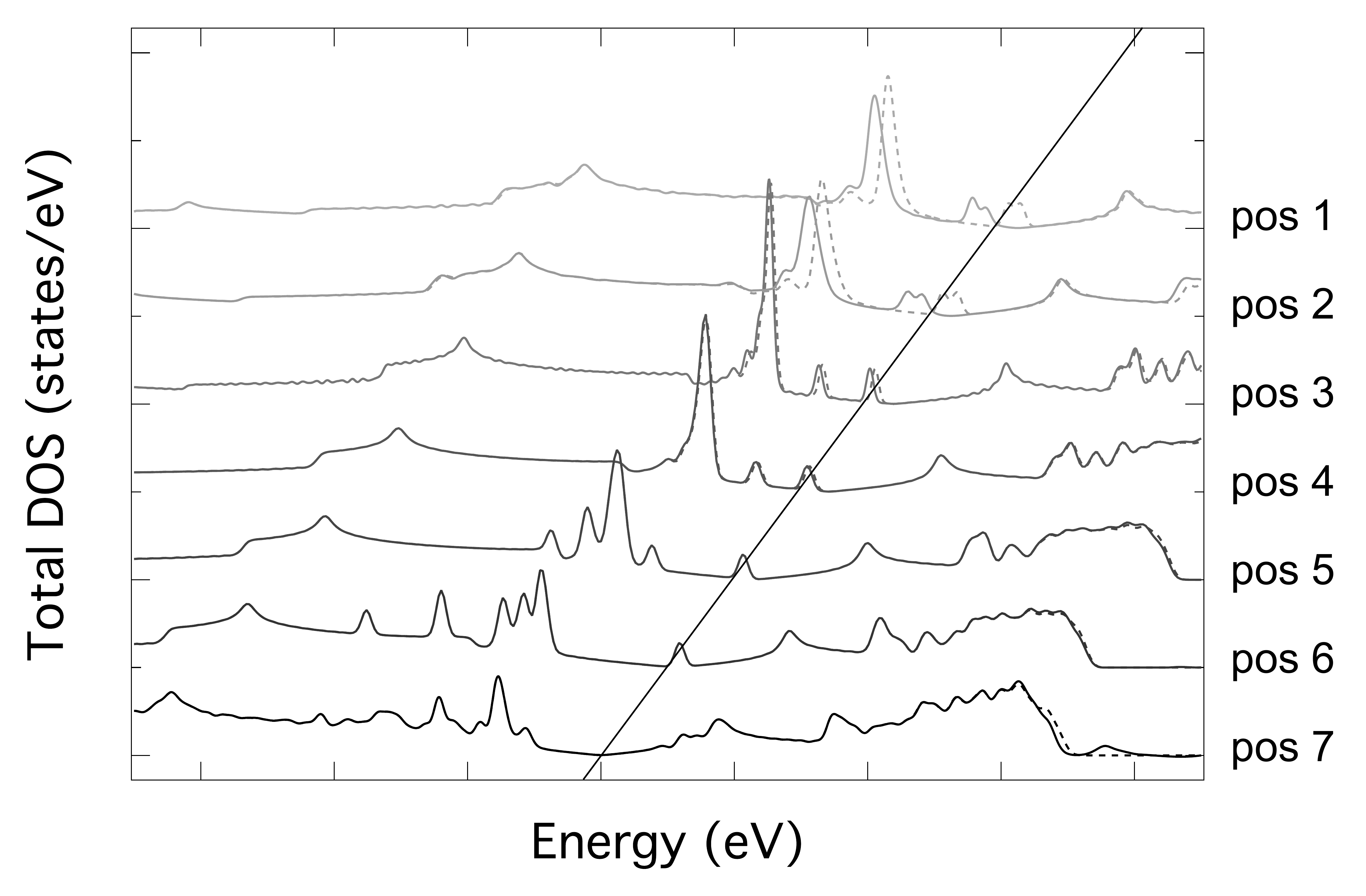}  
 \caption{Changes in total spin-polarized DOS (solid lines spin up; dashed lines spin down) when two pre-adsorbed adatoms come stepwise closer to each other to form Au$_{2}$/graphene, cf. Fig. \ref{fgr:Figure10}. The diagonal line marks the Fermi energies at the different positions.}
\label{fgr:Figure11}
\end{figure}

\begin{figure}[hbt]
\includegraphics[width=0.95\linewidth]{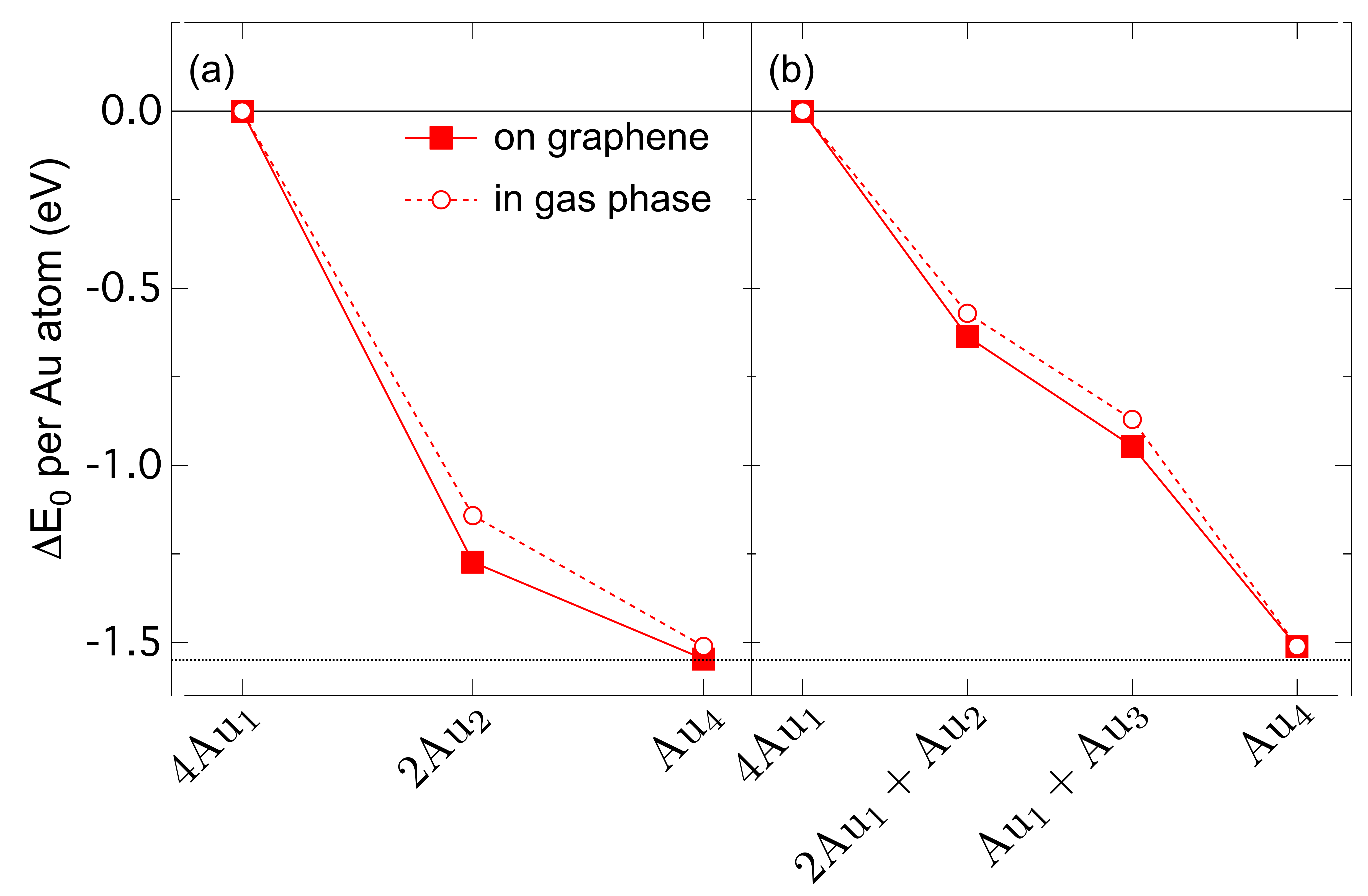}  
 \caption{(Color online) Changes in the total energy $\Delta E_{0}$ during the formation of Au$_{4}^{Y}$/graphene out of (a) two pre-adsorbed dimers, and (b) Au$_{4}^{D}$ out of four adatoms (solid squares). For comparison $\Delta E_{0}$ for the formation of Au$_{4}^{D}$ in the gas phase is shown as well (open circles). The dotted horizontal line marks the highest gain in energy.}
\label{fgr:Figure12}
\end{figure}

We modeled the initial clustering processes of gold on graphene in two steps. 
First, we simulated the formation of a dimer out of two pre-adsorbed gold atoms, cf. Figs \ref{fgr:Figure10} and \ref{fgr:Figure11}.
Second, two possible pathways for forming a tetramer, cf. Fig. \ref{fgr:Figure12}, were explored.
At the same time, the formation of a trimer out of a dimer and adatom was studied, too.

In order to simulate the clustering of two pre-adsorbed gold atom, they were initially placed as far away from each other as possible on a 5$\times$5 graphene sheet, i.e. 7.1~\AA.
This configuration is named position 1 in Figs \ref{fgr:Figure10} and \ref{fgr:Figure11}.
Since the total energy of the system decreases only marginally between position 1 and 2, we omitted intermediate steps, cf.  Fig. \ref{fgr:Figure10}.

In the subsequent steps, i.e. position 2 to 6, the second gold atom was brought closer to the fixed one along two C-C bonds, see inset in Fig. \ref{fgr:Figure10}.
For each of the positions 1 - 6 only the x-y-coordinates of the adatoms and one carbon atom at the rim were kept fixed, while z-coordinates of the adatoms and the rest of the graphene sheet could fully relax. 
In a final step we allowed the system to freely relax from position 6 and without an additional activation barrier it converged to the dimer ground state (position 7).
This final structure is identical to the one we found in Sec. \ref{sec:GS} by soft-landing the dimer on graphene, see Fig. \ref{fgr:Figure1} (b).

Both, in the gas phase and on graphene, the formation of a gold dimer is exothermic with a gain in total energy per gold atom of $\Delta E_{0}$ =  -1.14 and -1.29 eV, respectively, see Fig. \ref{fgr:Figure10}.
The higher energy gain on graphene is a result of the higher adsorption energy of Au$_{2}$ compared to two adatoms, see $E_{\mathrm{ads}}$ in Fig. \ref{fgr:Figure2} (a).
$\Delta E_{0}$ largely exceeds both the adsorption energy as well as the diffusion barriers of Au$_{1}$/graphene.
Thus the driving forces behind the clustering of two pre-adsorbed gold adatoms is the opportunity to lower the total energy of the system by forming strong gold-gold bonds, i.e. the strongly hybridized $d$-states in Fig. \ref{fgr:Figure6}.
At the positions 3 to 6 forces act on the two adatoms that try to pull them towards each other. 
This attractive, graphene mediated interaction has a maximal range of approximately 4.3~\AA, i.e. the Au-Au distance in position 3.
At this distance the two gold atoms are still well separated in the gas phase.

Note that the last two positions represent a dimer that is parallel (position 6) and perpendicular (position 7) to the graphene sheet. 
Obviously, in the gas phase the total energies have to be identical for position 6 and 7.
 
Figure \ref{fgr:Figure11} shows the changes in the spin-projected total DOS during the intermediate steps of forming a gold dimer on graphene, cf. position 1 to 7 in Fig. \ref{fgr:Figure10}.
Note that even for the largest possible separation of the two adatoms in the 5$\times$5 super-cell, i.e. 7.1~\AA\, (position 1), there is still a electrostatic interaction between the charged gold atoms, cf. Fig. \ref{fgr:Figure1} (a).
This can be seen from their 6$s$-states near the Fermi level that should exactly coincide without any mutual interaction.
At a Au-Au distance of 4.9~\AA\, (position 2) the 6$s$ peak splitting becomes more visible.
But still the 6$s$-states below the Fermi level belong solely to the spin-up channel.
When the graphene mediated attraction between the adatoms becomes obvious at a distance of 4.3 ~\AA, see $\Delta E_{0}$ at position 3 in Fig. \ref{fgr:Figure10}, one of the 6$s$ spins flips, leaving the total system with no total magnetic moments, as we saw above for the closed shell system Au$_{2}$/graphene.
Upon further approach of the two gold atoms the two atomic 6$s$-states form a molecular orbital that moves to lower energies.
Finally, it hybridizes with the $d_{z}^{2}$-states, cf. peaks 4 and 6 in Fig. \ref{fgr:Figure6}, in the dimer ground state structure (position 7).
Meanwhile the unoccupied 6$s$-states directly above the Fermi level in pos. 3 - 6 move up in energy to finally form peak no. 7 in Fig. \ref{fgr:Figure6}.

Our next aim was to study the formation of Au$_{3,4}$ on graphene.
We accomplished this by allowing two pre-adsorbed fragments to fully relax from an initial state corresponding to position 6 in Fig. \ref{fgr:Figure10}.
There are two different reaction pathways for forming a tetramer, resulting in the two isomers Au$_{4}^{D}$ and Au$_{4}^{Y}$, see Figs \ref{fgr:Figure1} (d) and (e).
In the first reaction pathway two dimers cluster from their perpendicular ground state into the Y shaped tetramer, 2 $\cdot$ Au$_{2}$ $\rightarrow$ Au$_{4}^{Y}$.
The energy gain per gold atom during this step is $\Delta E_{0}$ = -0.27 eV, see Fig. \ref{fgr:Figure12} (a).
Nonetheless, per dimer this energy gain is still bigger than the dimer adsorption energy of -0.47 eV, see Fig. \ref{fgr:Figure2} (a), overcompensating the breaking of one Au-C bond to form Au$_{4}^{Y}$/graphene.

The second pathway includes two intermediate steps to form Au$_{4}^{D}$ out of four single gold atoms.
To begin with, a trimer is formed, Au$_{1}$ + Au$_{2}$ $\rightarrow$ Au$_{3}$. 
Then, by adding yet an adatom, one forms the tetramer: Au$_{1}$ + Au$_{3}$ $\rightarrow$ Au$_{4}^{D}$.
The energy gain $\Delta E_{0}$ during the whole reaction is shown in Fig. \ref{fgr:Figure12} (b).
This process is exothermic and the fragments only have to overcome their diffusion barriers, Fig. \ref{fgr:Figure9}.
As for the dimer formation, the energy released by forming the strong Au-Au bonds is the driving force behind the clustering, which is for all steps bigger than the cost of breaking a Au-C bond or overcoming diffusion barriers. 

Although we did not explicitly calculate the stepwise movement of the different fragments, as we did in the case of two adatoms, see Fig. \ref{fgr:Figure10}, we expect a similar behavior due to the high mobility of the pre-adsorbed clusters, see Sec. \ref{sec:mobility}.

A comparison of $\Delta E_{0}$ in Figs \ref{fgr:Figure12} (a) and (b) shows, as mentioned in Sec. \ref{sec:GS}, that Au$_{4}^{Y}$/graphene is 137 meV lower in energy than Au$_{4}^{D}$/graphene.
The open symbols in Fig. \ref{fgr:Figure12} show the energies of the reaction steps in the gas phase.
Since Au$_{4}^{D}$ is the gas phase ground state structure, we assumed that both reaction pathways give this structure.
The end points in Fig. \ref{fgr:Figure12} (b) seem to coincide, i.e. the energy gain by clustering in the gas phase and on graphene appears to be the same.
That is actually the case since the adsorption energy of four Au$_{1}$ on graphene is -0.4 eV, which is almost identical to $E_{\mathrm{ads}}$ of Au$_{4}^{D}$ with -0.41 eV.

\section{\label{sec:summ}Summary and conclusions}
By means of density functional theory, the adsorption of Au$_{1-4}$ clusters on graphene, their mobility and clustering from pre-adsorbed fragments have been studied.
This allowed us to explain the initial steps in the experimentally observed formation of bigger gold aggregates on graphene.
While Au$_{1-3}$ adsorb similarly on graphene as on graphite, the tetramer turns out to be more peculiar.
In the gas phase it has two isomers, the diamond shaped Au$_{4}^{D}$, being the ground state, and the Y shaped Au$_{4}^{Y}$.
On graphene Au$_{4}^{Y}$ becomes the ground state instead.
Although Au$_{4}^{Y}$ binds stronger to the substrate and also receives more charge from it than Au$_{4}^{D}$ does, it disturbs the graphene sheet substantially less.

The cluster adsorption energies on graphene of all studied clusters range from -0.1 to -0.59 eV and substantially exceed their diffusion barriers, which are 4 to 36 meV, only.
All clusters diffuse along the C-C bonds, since none of them binds to the hollow sites in the centers of the carbon hexagons.

Our detailed analysis of the densities of states shows which states contribute to the strong intracluster bonds, predominantly the 5$d$-states, and which to the chemical bond between gold and carbon, i.e. mostly $s$-$d_{z^{2}}$-cluster states hybridizing with $p_{z}$-states of carbon.

The formation of the strong Au-Au bonds is the driving force behind the tendency to form bigger clusters from smaller pre-adsorbed fragments.
We studied these processes in detail for the stepwise formation of a dimer out of two pre-adsorbed adatoms, including the evolution of the electronic structure during the clustering.
The exothermic reaction has no additional activation barriers apart from the small diffusion barriers of the fragments moving along the C-C bonds.
We find the graphene mediated, attractive interaction of two gold adatoms to have a range of approximately 4.3~\AA, which is clearly longer than in the gas phase. 

Finally, we have also studied the formation of the trimer and tetramers by the reactions Au$_{1}$ + Au$_{2}$ $\rightarrow$ Au$_{3}$, Au$_{1}$ + Au$_{3}$ $\rightarrow$ Au$_{4}^{D}$, and 2 $\cdot$ Au$_{2}$ $\rightarrow$ Au$_{4}^{Y}$.
Both reactions are also activation barrier free.
As for the dimer, the energy gained by clustering, i.e. forming Au-Au bonds, exceeds substantially both the Au-C bond energies as well as the diffusion barriers.
Note that both tetramer isomers, i.e. Au$_{4}^{Y,D}$ are formed on graphene, depending on the actual reaction pathway.

\begin{acknowledgments}
This research was supported by the Swedish Energy Agency (Energimyndigheten) and the Swedish Research Council (Vetenskapsr\aa det).
Computation time on the Neolith Cluster at the National Supercomputer Centre in Link\"oping and on the Akka Cluster at the High Performance Computing Center North in Ume\aa\, was granted by the Swedish National Infrastructure for Computing (SNIC).
O. E. is grateful to the European Research Council (ERC) for support.
\end{acknowledgments}

\bibliography{references}

\end{document}